\documentclass[conference]{IEEEtran}
\usepackage[top=0.75in, bottom=0.75in, left=0.65in, right=0.65in]{geometry}

\usepackage{amsmath,amssymb,amsfonts,amsthm}

\newtheorem{theorem}{Theorem}

\newtheorem{lemma}{Lemma}

\usepackage{lipsum}

\usepackage{cite}

\usepackage{algorithm}
\usepackage{algpseudocode}

\usepackage{xcolor}
\usepackage{colortbl}
\usepackage{graphicx}
\usepackage{bm}

\usepackage{subfigure}

\begin{document}

\title{\fontsize{22.7pt}{25pt}\selectfont 
UAV-Assisted Downlink Satellite Covert Communication
}



\author{
\IEEEauthorblockN{Hung D. Nguyen and Jeongseok Ha}
\IEEEauthorblockA{School of Electrical Engineering, Korea Advanced Institute of Science and Technology, Daejeon, South Korea 34141.\\
Email: hungnd@kaist.ac.kr, jsha@kaist.edu\\[-4pt]}
}

\maketitle

\begin{abstract}
    This paper investigates the use of an unmanned aerial vehicle (UAV) to assist covert communication between a low-Earth orbit (LEO) satellite and a ground user under the surveillance of a passive warden. The UAV simultaneously serves its own ground network and acts as a friendly jammer to enhance the covertness of satellite transmissions. We derive a closed-form lower bound on the warden's average minimum detection error probability which is then used to define the covert constraint. Building on this, we formulate an optimization problem to jointly design the UAV's 3D placement, its power allocation, and the satellite's transmit power to maximize the system's covert rate. To solve the resulting non-convex problem, we propose an algorithm based on the block coordinate descent (BCD) and successive convex approximation (SCA) techniques, and further develop a Dinkelbach's algorithm for a special case. Numerical results validate the tightness of the derived bound and demonstrate the effectiveness of the proposed algorithms in configuring optimal system parameters.
\end{abstract}

\vspace{-5pt}

\section{Introduction} 

\IEEEPARstart{I}{n} recent years, low-Earth orbit (LEO) satellites have played a pivotal role in reshaping sixth-generation (6G) network architectures, attracting significant attention from both academia and industry \cite{Yue2023Low,Kim2024Downlink, Talgat2024Stochastic}. LEO satellite constellations such as Starlink, Telesat, and OneWeb are rapidly expanding, bringing the vision of seamless global connectivity with low latency and high data rates closer to reality. However, the wide coverage and inherent broadcast nature of LEO satellite communication also introduce critical security challenges that must be addressed to ensure the security and resilience of future 6G networks \cite{Yue2023Low}.

Covert communication, which aims to hide the very existence of information transmission from potential adversaries, has emerged as a promising paradigm for safeguarding sensitive information in future communication systems \cite{Chen2023Covert}. By exploiting inherent wireless channel uncertainties such as noise and fading, and/or introducing artificial uncertainties through techniques like power randomization and jamming, covert communication can effectively conceal transmissions from detection \cite{Chen2023Covert}. Compared to conventional encryption, it offers a higher level of security, guided by a simple yet profound principle: what cannot be seen cannot be attacked \cite{Kim2025Covert}.

Despite the potential of covert communication to enhance the security of satellite communication networks, research on satellite covert communication is still in its early stages. The authors in \cite{Song2023RIS} investigated the use of reconfigurable intelligent surfaces (RIS) to achieve covert communication in satellite downlink systems. In \cite{Zhang24Covert}, covert communication is enabled by leveraging random sidelobe interference from neighboring satellites to mask the covert signal at the warden. A related study \cite{Yu2025Covert} proposed a non-orthogonal multiple access (NOMA)-based covert communication scheme, where the satellite randomizes its transmit power to confuse the warden while simultaneously serving both a public user and a covert user.

Unmanned aerial vehicles (UAVs) have been widely recognized as valuable aerial complements to terrestrial and satellite networks in 6G, owing to their low cost and flexible deployment \cite{Yang2018Joint, Hourani2014Optimal}. However, the use of UAVs to assist satellite covert communication remains largely unexplored. To the best of our knowledge, only one recent work \cite{Mu2025Finite} has considered this problem, wherein a friendly UAV jammer aids the covert transmission of a satellite-UAV relay-ground link in the presence of a UAV warden. The work, however, focused on performance analysis in the finite blocklength regime, without optimizing key system parameters such as UAV placement and power allocation. Furthermore, the aforementioned works \cite{Song2023RIS, Zhang24Covert, Yu2025Covert, Mu2025Finite} assumed that the instantaneous channel state information (CSI) of the warden is available to the legitimate parties, and base their covertness analysis and/or optimization on the warden's detection error probability (DEP) viewpoint. This assumption is often overly optimistic in practice, since the warden's CSI is typically unavailable \cite{Shahzad2018Achieving}.

Motivated by the research gap and limitations mentioned above, this paper investigates UAV-assisted downlink satellite covert communication. Assuming the warden is passive and that only statistical CSI of the warden is available to the legitimate parties, we derive a closed-form lower bound on the warden's average minimum DEP. In addition, we jointly optimize the UAV's 3D placement, its transmit power allocation, and the satellite's transmit power to maximize the system's covert rate. To tackle the formulated non-convex problem, we employ a solution framework that integrates the block coordinate descent (BCD) with successive convex approximation (SCA) techniques. We also propose a Dinkelbach's algorithm for a special case of the problem. The tightness of the derived lower bound and the effectiveness of the proposed algorithms are confirmed through numerical experiments.

\section{System Model} \label{sec.SysModel}\vspace{-5pt}

\subsection{Scenario}

\begin{figure}[t] 
    \centering
    \includegraphics[width=0.75\linewidth]{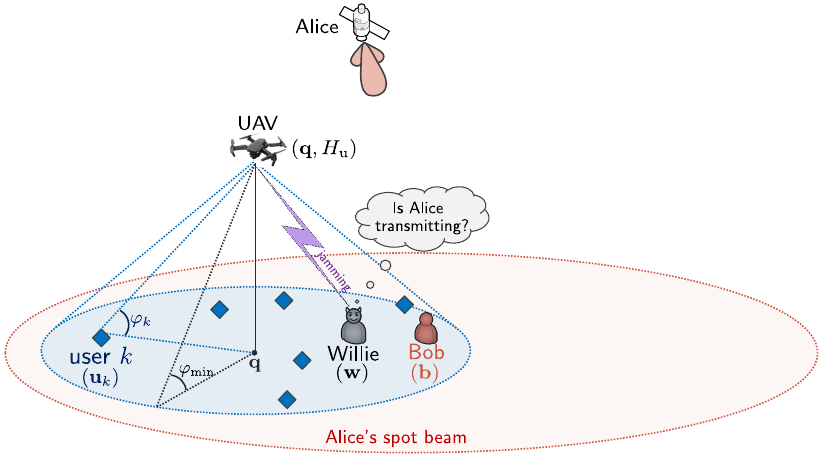}
    \caption{UAV-assisted downlink satellite covert communication network.}
    \label{Fig:Model}
\end{figure}

As illustrated in Fig.~\ref{Fig:Model}, we consider a downlink transmission from an LEO satellite (Alice) to a covert user (Bob) under the surveillance of a warden (Willie). Alice is equipped with multiple phased-array or dish antennas to form multiple spot beams on the ground \cite{Kim2024Downlink}. Bob and Willie, each equipped with a single antenna, are located within the same spot beam, with their ground coordinates denoted as $\mathbf{b} = [x_{\rm b}, y_{\rm b}]^T$ and $\mathbf{w} = [x_{\rm w}, y_{\rm w}]^T$, respectively. Owing to their close proximity within the beam, Bob and Willie experience highly similar received signal strengths, making covert communication particularly challenging for Alice. To overcome this, Alice enlists the assistance of a UAV operating near Bob and Willie to act as a friendly jammer, thereby aiding her covert transmission.

In this work, we assume that the UAV is equipped with a single antenna and serves its own network consisting of $K$ single-antenna user equipments (UEs). It employs frequency division multiple access to serve the UEs, allocating each UE a dedicated frequency band for communication. The frequency band of the UAV network does not overlap with that of the satellite network. Thus, the UAV's service to its UEs does not interfere with the communication between Alice and Bob, and vice versa. We denote the ground-projected coordinate and the altitude of the UAV by $\mathbf{q} = [x_{\rm u}, y_{\rm u}]^T$ and $H_{\rm u}$, respectively. Besides, the ground coordinates of the $K$ UEs are represented by $\mathbf{u}_k = [x_k, y_k]^T$, where $k \in \mathcal{K} \triangleq \{1, \dots, K\}$.

To assist Alice's covert transmission, the UAV transmits jamming signals in the Alice's frequency band to confuse Willie in his detection. The jamming power $P_{\rm j}$ follows a uniform distribution \cite{Shahzad2018Achieving}, i.e., $P_{\rm j} \sim \mathcal{U}(0, \hat{P}_{\rm j})$, where $\hat{P}_{\rm j}$ denotes the maximum power the UAV can allocate for jamming. We assume that the UAV and Bob have pre-shared secret key that allows Bob to cancel the UAV's jamming signal. However, the jamming signal may not be perfectly canceled at Bob due to practical hardware limitations, which results in residual interference \cite{Sena20Massive}. 
It is also assumed that the UAV assists Alice's covert transmission only when it has residual power after serving its UEs.

\subsection{Communication Model}
\subsubsection{Satellite Channel Model}
The channel gain from Alice to Bob/Willie can be expressed as
\begin{equation}
    g_{\rm a\chi} = \ell_{\rm a\chi} |h_{\rm a\chi}|^2,  \qquad \chi \in \{ \rm b,w\},
\end{equation}
where $\ell_{\rm a\chi} = [c/(4\pi f_{\rm c} d_{\rm a\chi})]^2 G_{\rm a\chi}$ represents the large-scale channel gain accounting for the free-space pathloss and the product of Alice's transmit and Bob/Willie's receive antenna gains $G_{a\chi}$. Here, $c$ is the speed of light, $f_{\rm c}$ is the carrier frequency, and $d_{a\chi}$ is the distance between Alice and Bob/Willie. 

For small-scale fading $h_{\rm a\chi}$, we assume that Bob and Willie experience independent quasi-static shadowed-Rician (SR) fading channels which is a widely accepted model for satellite communication \cite{Kim2024Downlink, Talgat2024Stochastic}. As such, the small-scale fading gains, $|h_{\rm ab}|^2$ and $|h_{\rm aw}|^2$, are independent and identically distributed random variables following the squared SR distribution \cite{Kim2024Downlink} characterized by parameters $b$, $m$, and $\Omega$. 
The cumulative distribution function (CDF) of $|h_{\rm a\chi}|^2$ is given by \cite{Talgat2024Stochastic}
\begin{equation}
    F_{|h_{\rm a\chi}|^2}(x)\!=\!\left(\!\frac{2bm}{2bm + \Omega}\!\right)^m \! \sum_{n=0}^{\infty} \frac{(m)_n \psi^n (2b)^{n}}{(n!)^2} \gamma\left(1+n, \frac{x}{2b}\right), 
    \label{eq.SR_CDF}
\end{equation}
where $\psi \triangleq \Omega/(2bm + \Omega)/(2b)$, $\Gamma(\cdot)$ is the Gamma function, $(x)_n = \Gamma(x + n)/\Gamma(x)$, and $\gamma(a, x) = \int_0^x t^{a-1} \exp(-t) dt$ represents the lower incomplete Gamma function. 

Due to the close proximity of Bob and Willie, their large-scale channel gains are assumed to be approximately equal, i.e., $\ell_{\rm ab} \approx \ell_{\rm aw}$, which is known to Alice from her knowledge of Bob's location. Also, Alice is assumed to know Bob's CSI perfectly but only the statistics of Willie's CSI, as Willie is a passive warden who does not transmit any signals.

\subsubsection{UAV Channel Model}
We assume the channel between each ground node and the UAV is dominated by a line-of-sight (LoS) path \cite{Yang2018Joint}. Accordingly, the channel gains from the UAV to the UEs and to Bob/Willie are respectively given by
\begin{align}
    g_{{\rm u} k} &= {\beta_0^{\rm \kappa}}/\left({\|\mathbf{q}-\mathbf{u}_k\|^2+H_{\rm u}^2}\right), \quad k \in \mathcal{K},  \\
    g_{{\rm u} \chi} &= {\beta_0^{\rm \chi}}/\left({\|\mathbf{q}-\bm{\chi}\|^2+H_{\rm u}^2}\right), \quad \bm{\chi} \in \{ \mathbf{b}, \mathbf{w} \},
\end{align}
where $\beta_0^\kappa$ and $\beta_0^\chi$ denote the channel power at a reference distance $d_0 = 1$~m for the UEs and for Bob/Willie, respectively. Note that $\beta_0^\kappa$ and $\beta_0^\chi$ are generally different since the UEs and Bob/Willie operate in different frequency bands.  
According to \cite{Hourani2014Optimal}, an LoS channel is highly likely when the elevation angle from a ground node to the UAV is sufficiently large. Accordingly, in this work, we require the UAV to maintain an elevation angle of at least $\varphi_{\rm min}$ with respect to every ground nodes (as depicted in Fig.~\ref{Fig:Model}) to reduce the risk of blockage which could degrade the UAV's service to UEs, weaken the jamming impact on Willie, or compromise the secret-key exchange with Bob.
This requirement leads to the following geometric constraints on the UAV's hovering position:
\begin{align} 
    \|\mathbf{q} - \mathbf{u}_k\| &\leq H_{\rm u}/\tan(\varphi_{\rm min}), \quad k \in \mathcal{K}, \label{eq.users_coor_constr} \\
    \|\mathbf{q} - \bm{\chi}\| &\leq H_{\rm u}/\tan(\varphi_{\rm min}), \quad \bm{\chi} \in \{ \mathbf{b}, \mathbf{w} \}. \label{eq.willie_coor_constr}
\end{align}
To guarantee \eqref{eq.users_coor_constr} and \eqref{eq.willie_coor_constr}, we assume the locations of all ground nodes are known by the UAV. Note that although Willie operates in a passive manner, the UAV can still determine his location by using onboard optical cameras and synthetic aperture radar with advanced image processing and pattern recognition techniques \cite{Yao2025UAV}.

\subsubsection{Achievable Rates} 
Based on the above system and channel models, the achievable rate (in bit/s/Hz) of UE $k$ ($k \in \mathcal{K}$) is given by $R_k = \log_2(1+{P_k g_{{\rm u}k}}/{\sigma_\kappa^2})$, where $P_k$ denotes the allocated power to UE $k$ and $\sigma_\kappa^2$ is the UE's noise variance which is identical among UEs under equal bandwidth allocation. On the other hand, the achievable covert rate of Alice's transmission to Bob (in bit/s/Hz) is expressed as
\begin{equation} \label{eq.R_b}
    R_{\rm b} = \log_2\left(1+\frac{P_{\rm a} g_{\rm ab}}{\varpi P_{\rm j} g_{\rm ub} +  \sigma_{\rm b}^2}\right),
\end{equation}
where $P_{\rm a}$ denote Alice's transmit power, $\varpi\!\in\![0,1]$ is Bob's interference cancellation coefficient, and $\sigma_{\rm b}^2$ is the noise variance at Bob.

\section{Covert Performance Analysis} \label{sec.Covertness}
\subsection{Detection at Willie}
The detection of Alice's transmission at Willie can be formulated as a binary hypothesis testing problem. Under the null hypothesis $\mathcal{H}_0$, Alice remains silent, whereas under the alternative hypothesis $\mathcal{H}_1$, Alice transmits covert signals. Accordingly, the received signal at Willie in the $n$-th channel use, $n = 1,\dots, N$, under $\mathcal{H}_0$ and $\mathcal{H}_1$, can be expressed as
\begin{equation*}
    y_{\rm w}[n]=\begin{cases}
        \sqrt{P_{\rm j}g_{\rm u w}}\, x_{\rm j}[n] + n_{\rm w}[n], & \mathcal{H}_0, \\[2pt]
        \sqrt{P_{\rm a} \ell_{\rm aw}}\, h_{\rm aw} x_{\rm a}[n]+\sqrt{P_{\rm j} g_{\rm u w}}\, x_{\rm j}[n]+n_{\rm w}[n], & \mathcal{H}_1,
    \end{cases}
\end{equation*}
where $x_{\rm a}[n] \sim \mathcal{CN}(0,1)$ and $x_{\rm j}[n] \sim \mathcal{CN}(0,1)$ denote the Alice's covert signal and the UAV's jamming signal, respectively. In addition, $n_{\rm w}[n] \sim \mathcal{CN}(0,\sigma_{\rm w}^2)$ is the additive white Gaussian noise (AWGN) at Willie, where $\sigma_{\rm w}^2$ is the Willie's noise variance.

Assuming that Willie adopts an energy detector, with sufficiently large block length, i.e., $N \to \infty$, the average received power at Willie can be written as
\begin{equation*}
    T = \lim_{N\to\infty}\frac{1}{N}\sum_{n=1}^{N}\left|y_{\rm w}[n]\right|^2
     = \begin{cases}
            P_{\rm j} g_{\rm u w}+\sigma_{\rm w}^2,  &\mathcal{H}_0, \\
            P_{\rm a} g_{\rm a w}+P_{\rm j} g_{\rm u w}+\sigma_{\rm w}^2, &\mathcal{H}_1.
        \end{cases}
\end{equation*}
Here, note that although Willie knows the distribution of the UAV's jamming power, i.e., $P_{\rm j} \sim \mathcal{U}(0, \hat{P}_{\rm j})$, the actual realization of $P_{\rm j}$ within a given block is unknown to him. Based on the test statistic $T$, Willie makes a binary decision on whether Alice transmits ($\mathcal{D}_1$) or not ($\mathcal{D}_0$) by performing the hypothesis testing $T \gtrless_{\mathcal{D}_0}^{\mathcal{D}_1} \tau$, where $\tau$ is the detection threshold. In this process, Willie inevitably encounters two types of decision errors namely  1) false alarm (FA), when $\mathcal{D}_1$ is chosen while $\mathcal{H}_0$ is true, and 2) missed detection (MD), when $\mathcal{D}_0$ is chosen while $\mathcal{H}_1$ is true. 
The probabilities of FA and MD depend on the Willie's selection of threshold $\tau$.

\begin{lemma}
    For a given detection threshold $\tau$, the FA and MD probabilities at Willie are respectively given by\vspace{-5pt}
    \begin{align}
    \mathbb{P}_{\rm FA}
         &= \begin{cases}
            1, & \tau < \rho_1, \\[-3pt]
            1-\frac{\tau-\sigma_{\rm w}^2}{\hat{P}_{\rm j} g_{\rm u w}},  \quad &\rho_1 \leq \tau < \rho_2, \\[-3pt]
            0, & \tau \geq \rho_2,
         \end{cases}\label{eq.FA} \\
    \mathbb{P}_{\rm MD}
         &= \begin{cases}
            0, & \tau < \rho_3, \\[-3pt]
            \frac{\tau-P_{\rm a}g_{\rm a w}-\sigma_{\rm w}^2}{\hat{P}_{\rm j} g_{\rm u w}}, & \rho_3 \leq \tau < \rho_4, \\[-3pt]
            1, & \tau \geq \rho_4,
         \end{cases}\label{eq.MD} \\[-20pt] \notag
    \end{align}
    where $\rho_1 \triangleq \sigma_{\rm w}^2$, $\rho_2 \triangleq \hat{P}_{\rm j} g_{\rm u w}+\sigma_{\rm w}^2$, $\rho_3 \triangleq  P_{\rm a}g_{\rm a w}+\sigma_{\rm w}^2$, and $\rho_4 \triangleq \hat{P}_{\rm j}g_{\rm u w}+$ $P_{\rm a}g_{\rm a w}+\sigma_{\rm w}^2$.
\end{lemma}
\begin{IEEEproof}
    The FA and MD probabilities are derived as\vspace{-2pt}
    \begin{align}
        \hspace{-5pt}\mathbb{P}_{\rm FA}& \triangleq \mathbb{P}(\mathcal{D}_1 \mid \mathcal{H}_0)=\mathbb{P}\left(P_{\rm j}g_{\rm u w}+\sigma_{\rm w}^2>\tau\right) \notag \\
        & =\mathbb{P}\left(P_{\rm j}>\frac{\tau-\sigma_{\rm w}^2}{g_{\rm u w}}\right)=1-F_{P_{\rm j}}\left(\frac{\tau-\sigma_{\rm w}^2}{g_{\rm u w}}\right), \label{eq.FA_derivation}\\
        \hspace{-5pt}\mathbb{P}_{\rm MD}
        & \triangleq \mathbb{P}(\mathcal{D}_0 \mid \mathcal{H}_1)
          =\mathbb{P}\left(P_{\rm a}g_{\rm a w}+P_{\rm j}g_{\rm u w}+\sigma_{\rm w}^2<\tau\right) \notag \\
        & =\mathbb{P}\!\left(\!P_{\rm j}\!<\!\frac{\tau\!-\!P_{\rm a}g_{\rm a w}\!-\!\sigma_{\rm w}^2}{g_{\rm u w}}\!\right)\!=\!F_{P_{\rm j}}\!\left(\!\frac{\!\tau\!-\!P_{\rm a}g_{\rm a w}\!-\!\sigma_{\rm w}^2}{g_{\rm u w}}\!\right)\!,\hspace{-5pt}\label{eq.MD_derivation} 
    \end{align} 
    where $F_{P_{\rm j}}(x)$ in \eqref{eq.FA_derivation} and \eqref{eq.MD_derivation} represents the CDF of $P_{\rm j}$. Since $P_{\rm j} \sim \mathcal{U}(0,\hat{P}_{\rm j})$, we have $F_{P_{\rm j}}(x) = 0$ for $x<0$, $F_{P_{\rm j}}(x) = x/\hat{P}_{\rm j}$ for $0\le x < \hat{P}_{\rm j}$, and $F_{P_{\rm j}}(x) = 1$ for $x\ge \hat{P}_{\rm j}$. Substituting this CDF into \eqref{eq.FA_derivation} and \eqref{eq.MD_derivation} and simplifying the expressions yields the FA and MD probabilities in \eqref{eq.FA} and \eqref{eq.MD}, respectively.
\end{IEEEproof}

Assuming Willie is unaware of when Alice will transmit, and Alice transmits with equal \emph{a priori} probabilities, i.e., $\mathbb{P}(\mathcal{H}_0)\!=\!\mathbb{P}(\mathcal{H}_1)\!=\!\frac{1}{2}$, the DEP at Willie is derived as 
\begin{equation}
    \xi = \mathbb{P}_{\rm FA} + \mathbb{P}_{\rm MD}.
    \label{eq.DEP_def}
\end{equation}
Willie aims to minimize his DEP by optimally selecting $\tau$.
\begin{lemma}
    Willie's detection threshold can be optimally set to $\tau^\star = \rho_2$, which yields a minimum DEP of
    \begin{equation}
        \xi^\star =
        \begin{cases}
            1-\frac{P_{\rm a}g_{\rm a w}}{\hat{P}_{\rm j}g_{\rm u w}}, & \text{if } \rho_2 > \rho_3, \\
            0, & \text{if } \rho_2 \leq \rho_3.
        \end{cases}
        \label{eq.min_DEP}   
    \end{equation}
\end{lemma}
\begin{IEEEproof}
    From the definitions of $\rho_1, \rho_2, \rho_3$, and $\rho_4$, we have $\rho_1\!<\!\rho_2\! <\rho_4$ and $\rho_1\!<\rho_3\!<\!\rho_4$, while the ordering between $\rho_2$ and $\rho_3$ is uncertain. Thus, two cases are considered.
    \begin{itemize}
        \item If $\rho_2 \leq \rho_3$, Willie simply picks an arbitrary  $\tau \in [\rho_2, \rho_3]$, which gives him $\mathbb{P}_{\mathrm{FA}}=\mathbb{P}_{\mathrm{MD}} =0$ and thus $\xi^\star = 0$.
        \item If $\rho_2 > \rho_3$, substituting \eqref{eq.FA} and \eqref{eq.MD} into \eqref{eq.DEP_def} gives \vspace{-5pt}
        \begin{equation}
        \xi = 
            \begin{cases}
                1, & \tau < \rho_1, \\[-2pt]
                1-\frac{\tau-\sigma_{\rm w}^2}{\hat{P}_{\rm j} g_{\rm u w}}, & \rho_1 \leq \tau < \rho_3, \\
                1-\frac{P_{\rm a}g_{\rm a w}}{\hat{P}_{\rm j}g_{\rm u w}}, & \rho_3 \leq \tau < \rho_2,\\
                \frac{\tau-P_{\rm a}g_{\rm a w}-\sigma_{\rm w}^2}{\hat{P}_{\rm j} g_{\rm u w}}, & \rho_2 \leq \tau < \rho_4,\\[-2pt]
                1, & \tau \geq \rho_4.
            \end{cases}\vspace{-3pt}
            \label{eq.DEP_2ndcase}
        \end{equation}
        From \eqref{eq.DEP_2ndcase}, it can be shown that $\xi$ is minimized when $\tau \in [\rho_3, \rho_2]$. Thus, Willie chooses $\tau \in [\rho_3, \rho_2]$, which gives him a minimum DEP of $\xi^\star = 1-({P_{\rm a}g_{\rm a w}})/({\hat{P}_{\rm j}g_{\rm u w}})$.
    \end{itemize}
    In both cases, selecting $\tau^\star = \rho_2$ is sufficient for Willie to achieve the minimum DEP. The proof is completed.
\end{IEEEproof}

It should be noted that \eqref{eq.min_DEP} is the minimum DEP from Willie's perspective, assuming the worst-case scenario where he has perfect knowledge of $\sigma_{\rm w}^2, g_{\rm aw}$, $g_{\rm uw}$, $P_{\rm a}$, and $\hat{P}_{\rm j}$.

\subsection{Covert Constraint Analysis}
To guarantee covertness, Alice aims to ensure that the minimum DEP at Willie satisfies: $\xi^\star \geq 1-\varepsilon$, where $0 \leq \varepsilon < 0.5$ is a small value that quantifies the covertness level \cite{Chen2023Covert}. However, since $\xi^\star$ in \eqref{eq.min_DEP} depends on the random small-scale fading channel gain $|h_{\rm a w}|^2$, which is unknown to Alice, she cannot directly enforce this covert constraint. To address this, Alice considers the \emph{average} minimum DEP over the distribution of $|h_{\rm aw}|^2$, denoted by $\bar{\xi}^\star$, and instead imposes the constraint $\bar{\xi}^\star \geq 1-\varepsilon$. From the expression of $\xi^\star$ in \eqref{eq.min_DEP} and by using the law of total expectation, $\bar{\xi}^\star$ can be calculated as follows:
\begin{align}
    \bar{\xi}^\star 
    &\triangleq  \mathbb{E}_{|h_{\rm aw}|^2}[\xi^\star] = \mathbb{E}_{|h_{\rm aw}|^2}\left[\xi^\star|\rho_2\!>\!\rho_3\right]\!+\!\underbrace{\mathbb{E}_{|h_{\rm aw}|^2}\left[\xi^\star|\rho_2 \leq  \rho_3\right]}_{= 0} \notag \\[-6pt]
    &= \int_0^{\frac{\hat{P}_{\rm j} g_{\rm u w}}{P_{\rm a} \ell_{\rm aw}}} \left( 1-\frac{P_{\rm a }\ell_{\rm aw} x}{\hat{P}_{\rm j}g_{\rm u w}} \right) f_{|h_{\rm aw}|^2}(x)\,dx \notag \\[-1pt]
    &= F_{|h_{\rm aw}|^2}\!\left(\!\frac{\hat{P}_{\rm j} g_{\rm u w}}{P_{\rm a} \ell_{\rm aw}}\!\right) \!-\! \frac{P_{\rm a }\ell_{\rm aw}}{\hat{P}_{\rm j}g_{\rm u w}} \int_0^{\frac{\hat{P}_{\rm j} g_{\rm u w}}{P_{\rm a} \ell_{\rm aw}}}\! x f_{|h_{\rm aw}|^2}(x)\,dx, \label{eq.avg_min_DEP}
\end{align}
where $f_{|h_{\rm aw}|^2}(x)$ represents the probability density function (PDF) of $|h_{\rm aw}|^2$, and $F_{|h_{\rm aw}|^2}(x)$ is the CDF of $|h_{\rm aw}|^2$ given in \eqref{eq.SR_CDF}. The closed-form expression of $\bar{\xi}^\star$, though can be expressed in terms of summation and integration with special functions, is very complex and intractable for solving the optimization problem presented later. To overcome this challenge, we derive a tractable lower bound for $\bar{\xi}^\star$ in the following theorem.\vspace{-3pt}
\begin{theorem} \label{theorem.lower_bound}
    Willie's average minimum DEP can be lower bounded as \vspace{-4pt}
    \begin{equation}
        \bar{\xi}^\star > \bar{\xi}_{\rm lb}^\star \triangleq 1 - \alpha \exp \left(- \frac{\mu \hat{P}_{\rm j}g_{\rm uw}}{\theta P_{\rm a}\ell_{\rm aw}}\right) - \frac{\alpha \theta P_{\rm a}\ell_{\rm aw}}{\hat{P}_{\rm j}g_{\rm uw}},\vspace{-5pt}
        \label{eq.avg_min_DEP_lb}
    \end{equation}
    where\vspace{-3pt}
    \begin{equation}
        \alpha \triangleq \frac{m(2b + \Omega)^2}{4mb^2\!+\!4mb\Omega\!+\!\Omega^2}, \quad  \theta \triangleq \frac{4mb^2\!+\!4mb\Omega\!+\!\Omega^2}{m(2b + \Omega)},\vspace{-3pt}
        \label{eq.Gamma_params}
    \end{equation}
    and $\mu \triangleq (\alpha !)^{-1/\alpha}$, with $b$, $m$, and $\Omega$ being the SR parameters.
\end{theorem}
\begin{IEEEproof}
    We adopt the technique in \cite{Talgat2024Stochastic} and approximate the SR distribution with a Gamma distribution. Accordingly, the CDF of $|h_{\rm aw}|^2$ can be approximated as $F_{|h_{\rm aw}|^2}(x) \approx \gamma(  \alpha, x/ \theta )/\Gamma(\alpha)$, where $\alpha$ and $\theta$ are the shape and scale parameters of the Gamma distribution, respectively, which can be determined through the SR parameters as in \eqref{eq.Gamma_params} \cite{Talgat2024Stochastic}. 
    For $\alpha>1$, which holds for most of the shadowing levels as shown in Table~\ref{Tab.SR_Para}, the CDF of the Gamma distribution can be tightly lower bounded by $F_G(x) > (1 - \exp(-\mu x/\theta))^{\alpha}$, where $\mu \triangleq (\alpha !)^{-1/\alpha}$ \cite{Talgat2024Stochastic}. Using the first order Taylor expansion for small value of $\exp(-\mu x/\theta)$, we further have $(1 - \exp(-\mu x/\theta))^{\alpha} > 1 -\alpha \exp(-\mu x/\theta)$. As a result, the first term in \eqref{eq.avg_min_DEP} can be lower bounded as\vspace{-2pt}
    \begin{equation} \label{eq.CDF_lb}
       F_{|h_{\rm aw}|^2} \left( \frac{\hat{P}_{\rm j} g_{\rm u w}}{P_{\rm a} \ell_{\rm aw}} \right) > 1 - \alpha \exp \left(- \frac{\mu \hat{P}_{\rm j} g_{\rm u w}}{\theta P_{\rm a} \ell_{\rm aw}} \right).\vspace{-2pt}
    \end{equation}
    Besides, the integral term in \eqref{eq.avg_min_DEP} can be upper bounded as\vspace{-2pt}
    \begin{equation} \label{eq.int_ub}
        \int_0^{\frac{\hat{P}_{\rm j} g_{\rm u w}}{P_{\rm  a} \ell_{\rm aw}}} x f_{|h_{\rm  aw}|^2}(x)dx <  \int_0^{\infty} x f_{|h_{\rm aw}|^2}(x)dx  \overset{(i)}{=}  \alpha \theta,\vspace{-2pt}
    \end{equation}
    where $(i)$ follows from the expectation of the Gamma distribution, i.e., $\mathbb{E}[|h_{\rm aw}|^2] = \alpha \theta$.
    By substituting \eqref{eq.CDF_lb} and \eqref{eq.int_ub} into \eqref{eq.avg_min_DEP}, we obtain the lower bound of $\bar{\xi}^\star$ as in \eqref{eq.avg_min_DEP_lb}. 
\end{IEEEproof}
Now, by using $\bar{\xi}_{\rm lb}^\star$ for the covert constraint, i.e., $\bar{\xi}_{\rm lb}^\star \geq 1-\varepsilon$, we ensure a conservative guarantee of system covertness. The constraint $\bar{\xi}_{\rm lb}^\star \geq 1-\varepsilon$ can be equivalently expressed as\vspace{-3pt}
\begin{equation}
    \alpha \exp \left(- \frac{\mu \hat{P}_{\rm j} g_{\rm uw}}{\theta P_{\rm a}\ell_{\rm aw}}\right)
    + \frac{\alpha\theta P_{\rm a}\ell_{\rm aw}}{\hat{P}_{\rm j} g_{\rm uw}}
    \leq \varepsilon.\vspace{5pt}
    \label{eq.covert_cons_reform}
\end{equation}

\section{Covert Rate Maximization} \label{sec.Optimization}
\subsection{Problem Formulation}
With limited power budget, the UAV's operating location plays a critical role in system performance, as it must simultaneously support UE service, provide effective jamming at Willie, and mitigate interference at Bob. To this end, we aim to jointly optimize the UAV's 3D placement, its power allocation, and Alice's transmit power to maximize the system's covert rate. The optimization problem can thus be formulated as\vspace{-2pt}
\begin{align*}
    \textbf{(P1):} \quad \quad  
    &\max _{\{P_k\}_{k=1}^K, \hat{P}_{\rm j}, P_{\rm a}, \mathbf{q}, H_{\rm u}} R_{\rm b} \\[-1pt]
    \text{s.t.} \qquad  
    & \mathcal{C}1: \quad  \eqref{eq.covert_cons_reform}, \\[-2pt]
    & \mathcal{C}2: \quad R_k \geq R_{\rm tg}, \quad \forall k \in \mathcal{K}, \\[-3pt]  
    & \mathcal{C}3: \quad \hat{P}_{\rm j} + \sum_{k=1}^K P_k \leq P_{\rm tot},\\[-3pt] 
    & \mathcal{C}4: \quad \hat{P}_{\rm j} \geq 0, \quad P_k \geq 0,~\forall k\in\mathcal{K},\\[-1pt]
    & \mathcal{C}5: \quad 0 \leq P_{\rm a} \leq P_{\rm a}^{\rm max }, \\[-1pt]
    & \mathcal{C}6: \quad H_{\rm u}^{\rm min } \leq H_{\rm u} \leq H_{\rm u}^{\rm max }, \\[-1pt]
    & \mathcal{C}7: \quad \eqref{eq.users_coor_constr}, \eqref{eq.willie_coor_constr}. \\[-15pt]
\end{align*}
In \textbf{(P1)}, $\mathcal{C}1$ is the covert constraint, $\mathcal{C}2$ guarantees that each UE achieves at least the target rate $R_{\rm tg}$. Besides, $\mathcal{C}3$ and $\mathcal{C}4$ are the power constraints of the UAV, where $P_{\rm tot}$ denotes its total communication power. $\mathcal{C}5$ imposes a restriction on Alice's transmit power. $\mathcal{C}6$ limits the UAV's altitude within the regulatory range $[H_{\rm u}^{\rm min}, H_{\rm u}^{\rm max}]$, while $\mathcal{C}7$ ensures that all ground nodes are within the UAV's effective coverage area, as discussed earlier.
\textbf{(P1)} is difficult to solve directly due to the non-concave objective function, non-convex constraints $\mathcal{C}1$, $\mathcal{C}2$, as well as the coupling among the optimization variables.

\begin{table}[t]
    \centering
    \caption{Shadowed-Rician parameters for different shadowing levels \cite{Kim2024Downlink}, and the corresponding Gamma parameters.}
        \begin{tabular}{c c c c c c}
         \hline
        Shadowing level & $b$ & $m$ & $\Omega$ & $\alpha $ & $\theta$ \\
         \hline\\[-7pt]
        Light  & 0.158  & 19.4  & 1.29 & 2.577 & 0.623  \\ 
        Average  & 0.126 & 10.1 & 0.835 & 2.135 & 0.509  \\ 
        Heavy   & 0.063 & 1 & $8.97 \times 10^{-4}$ & 1 & 0.127 \\ 
         \hline
    \end{tabular}
    \label{Tab.SR_Para}
\end{table}

\subsection{Proposed Solution}
To tackle the coupling among the optimization variables in \textbf{(P1)}, we propose a BCD-based algorithm which alternatively solves the following two subproblems until convergence.
\subsubsection{Optimization of \{$\mathbf{q}, H_{\rm u}$\} with fixed \{$\{P_k\}_{k=1}^K, \hat{P}_{\rm j}, P_{\rm a}$\}}
Given the UAV's power allocation and Alice's transmit power, $R_{\rm b}$ in \eqref{eq.R_b} is an increasing function of the UAV-Bob distance, $\sqrt{\|\mathbf{q} - \mathbf{b}\|^2 + H_{\rm u}^2}$. Thus, we can express the UAV placement optimization subproblem as follows:
\begin{equation*}
    \textbf{(P2):} \qquad  
    \max_{\mathbf{q}, H_{\rm u}} \left\{\|\mathbf{q} - \mathbf{b}\|^2 + H_{\rm u}^2 \right\}
    \quad \text{s.t.} \quad \mathcal{C}1, \mathcal{C}2, \mathcal{C}6, \mathcal{C}7.
\end{equation*}
To solve \textbf{(P2)}, we first focus on the covert constraint $\mathcal{C}1$. For positive $\alpha$ and $\mu$ defined in Theorem~\ref{theorem.lower_bound}, let us define the following function:
\begin{equation}
    \Phi(x) \triangleq \alpha \exp (-\mu x^{-1}) + \alpha x, \quad x>0. 
    \label{eq.Phi_def}
\end{equation}
Then, we have $\partial \Phi/ \partial x = \alpha \exp(-\mu x^{-1}) \mu x^{-2} + \alpha > 0$ which implies that $\Phi(x)$ is strictly increasing in $x$. Note that the left-hand side of \eqref{eq.covert_cons_reform} can be expressed as $\Phi (d_{\rm uw}^2/\Lambda) $, where $d_{\rm uw}^2 \triangleq \|\mathbf{q} - \mathbf{w}\|^2 + H_{\rm u}^2$ and $\Lambda \triangleq {\hat{P}_{\rm j}\beta_0^\chi}/{(\theta P_{\rm a} \ell_{\rm aw})}>0$. Thus, if we denote $x = d_{\rm uw}^2/\Lambda$, then since $\partial \Phi / \partial d_{\rm uw}^2\!=\!(\partial \Phi / \partial x)(\partial x / \partial d_{\rm uw}^2)\!=\!(\partial \Phi / \partial x)/\Lambda >0$, we have that $\Phi(d_{\rm uw}^2/\Lambda)$ is strictly increasing in $d_{\rm uw}^2$. Therefore, the covert constraint $\mathcal{C}1$, given by $\Phi (d_{\rm uw}^2/\Lambda) \leq \varepsilon$, is equivalent to
\begin{equation}
    \mathcal{C}1': \quad  d_{\rm uw}^2= \|\mathbf{q} - \mathbf{w}\|^2 + H_{\rm u}^2 \leq \Lambda \Phi^{-1}(\varepsilon),
    \label{eq.C1_reform}
\end{equation}
where the value of $\Phi^{-1}(\varepsilon)$ can be easily found by a line search.

Now, letting $\bar{\mathbf{q}} \triangleq [\mathbf{q}^T, H_{\rm u}]^T$, $\bar{\mathbf{w}} \triangleq [\mathbf{w}^T, 0]^T$, $\bar{\mathbf{b}} \triangleq [\mathbf{b}^T, 0]^T$, $\bar{\mathbf{u}}_k \triangleq [\mathbf{u}_k^T, 0]^T$, $\mathbf{Q} \triangleq [\mathbf{I}_{2 \times 2} \quad \mathbf{0}_{2 \times 1}]$, $\mathbf{c} \triangleq [0,  0, {1}/{\tan \varphi_{\min}}]^T$, and $\mathbf{e} \triangleq [0,  0, 1]^T$, \textbf{(P2)} can be equivalently rewritten as
\begin{align*}
    \textbf{(P2$^\prime$):} \qquad   &\max_{\bar{\mathbf{q}}}  \|\bar{\mathbf{q}} - \bar{\mathbf{b}}\|^2 \\[-3pt]
    \text{s.t.} \qquad
     \overline{\mathcal{C}1'}:\quad  &\|\bar{\mathbf{q}} - \bar{\mathbf{w}}\| \leq \sqrt{\Lambda \Phi^{-1}(\varepsilon)}, \\[-2pt]
     \overline{\mathcal{C}2}:\quad  &\|\bar{\mathbf{q}} - \bar{\mathbf{u}}_k\| \leq \sqrt{{P_k}/{\Xi}}, \quad \forall k \in \mathcal{K}, \\
     \overline{\mathcal{C}6}:\quad  &H_{\min} \leq \mathbf{e}^T\bar{\mathbf{q}} \leq H_{\max}, \\
     \overline{\mathcal{C}7}:\quad  &\|\mathbf{Q}\bar{\mathbf{q}} - \mathbf{u}_k\| \leq \mathbf{c}^T\bar{\mathbf{q}}, \quad \forall k \in \mathcal{K}, \\
    &\|\mathbf{Q}\bar{\mathbf{q}} - \boldsymbol{\chi}\| \leq \mathbf{c}^T\bar{\mathbf{q}}, \quad \boldsymbol{\chi} \in \{\mathbf{b}, \mathbf{w}\},
\end{align*}
where $\Xi \triangleq {(2^{R_{\rm tg}}-1)\sigma_\kappa^2}/{{\beta_0^\kappa}} $. \textbf{(P2$^\prime$)} is non-convex due to its non-concave objective function. To tackle this issue, we apply the SCA technique to relax it. Specifically, we note that the objective function $\|\bar{\mathbf{q}} - \bar{\mathbf{b}}\|^2$ is convex in $\bar{\mathbf{q}}$. Thus, for given $\bar{\mathbf{q}}^{(l)}$ from the $l$-th SCA iteration, we can construct a linear surrogate function $\mathcal{O}_{\rm lb}(\bar{\mathbf{q}}|\bar{\mathbf{q}}^{(l)})$ that globally minorizes $\|\bar{\mathbf{q}} - \bar{\mathbf{b}}\|^2$ using the first-order Taylor expansion around $\bar{\mathbf{q}}^{(l)}$ as follows:
\begin{equation*}
    \|\bar{\mathbf{q}}\!-\!\bar{\mathbf{b}}\|^2\!\geq\! 
    \mathcal{O}_{\rm lb}(\bar{\mathbf{q}}|\bar{\mathbf{q}}^{(l)})\!\triangleq\!
    \|\bar{\mathbf{q}}^{(l)}\!-\!\bar{\mathbf{b}}\|^2 + 2(\bar{\mathbf{q}}^{(l)}\!-\!\bar{\mathbf{b}})^T(\bar{\mathbf{q}}\!-\!\bar{\mathbf{q}}^{(l)}).
\end{equation*}
Then, \textbf{(P2$^\prime$)} can be transformed into\vspace{-1pt}
\begin{equation*}
    \textbf{(P2$^{\prime\prime}$):} \qquad   \max_{\bar{\mathbf{q}}}  \mathcal{O}_{\rm lb}(\bar{\mathbf{q}}|\bar{\mathbf{q}}^{(l)}) \quad   
    \text{s.t.} \quad \overline{\mathcal{C}1'}, \overline{\mathcal{C}2}, \overline{\mathcal{C}6}, \overline{\mathcal{C}7},\\[-1pt]
\end{equation*}
which is a second-order cone program (SOCP) that can be efficiently solved using existing tools such as CVX \cite{Boyd_CVX}.

\begin{algorithm}[t] \caption{Proposed BCD-based Algorithm for Solving \textbf{(P1)}}
     \small
    \begin{algorithmic}[1]
        \Statex \textbf{Input:} $m, b, \Omega, c, f_{\rm c}, d_{\rm a\chi}, G_{\rm a\chi}, \beta_0^\chi, \beta_0^\kappa$, $\varphi_{\rm min}$, $\sigma_\kappa^2$, $\sigma_{\rm b}^2$, $\sigma_{\rm w}^2$, $\varpi$, $H_{\rm min}$, $H_{\rm max}$, $R_{\rm tg}$, $\varepsilon$, $P_{\rm tot}$, $P_{\rm a}^{\rm max}$, $K$, $\mathbf{b}$, $\mathbf{w}$, $\{\mathbf{u}_k\}_{k=1}^K$,$I_{\rm max}$, $\delta$;
        \Statex \textbf{Output:} $\{P_k^\star\}_{k=1}^K, \hat{P}^\star_{\rm j}, P^\star_{\rm a}, \mathbf{q}^\star, H^\star _{\rm u}$.
        \Statex \textbf{Initialize:} Iteration index $l=0$, $\{P_k^{(0)}\}_{k=1}^K$, $\hat{P}_{\rm j}^{(0)}$, and $P_{\rm a}^{(0)}$;
        \State Compute $\Phi^{-1}(\varepsilon)$, where $\Phi(\cdot)$ is defined in \eqref{eq.Phi_def};
        \Repeat
            \State Given $\{P_k^{(l)}\}_{k=1}^K$, $\hat{P}_{\rm j}^{(l)}$, $P_{\rm a}^{(l)}$,  solve \textbf{(P2$^{\prime\prime}$)} and obtain $\mathbf{q}^{(l+1)}$, $H_{\rm u}^{(l+1)}$;
            \State Given $\mathbf{q}^{(l+1)}$, $H_{\rm u}^{(l+1)}$, solve \textbf{(P3$'$)} and obtain $\{P_k^{(l+1)}\}_{k=1}^K$, $\hat{P}_{\rm j}^{(l+1)}$, $P_{\rm a}^{(l+1)}$;
            \State Update $l \gets l + 1$;
        \Until{$|R_{\rm b}^{(l)} - R_{\rm b}^{(l-1)}| < \delta$ or $l>I_{\rm max}$};
        \State Set $P_k^\star \gets P_k^{(l)},~\forall k \in \mathcal{K}$ , $\hat{P}^\star_{\rm j} \gets \hat{P}_{\rm j}^{(l)}$, $P^\star_{\rm a} \gets P_{\rm a}^{(l)}$, $\mathbf{q}^\star \gets \mathbf{q}^{(l)}$, and $H^\star _{\rm u} \gets H_{\rm u}^{(l)}$;\vspace{-1pt}
    \end{algorithmic}
    \label{Alg.BCD}
\end{algorithm}

\subsubsection{Optimization of \{$\{P_k\}_{k=1}^K, \hat{P}_{\rm j}, P_{\rm a}$\} with fixed \{$\mathbf{q}, H_{\rm u}$\}}
Given the UAV's placement, the power optimization subproblem can be formulated as follows:\vspace{-2pt}
\begin{equation*}
    \textbf{(P3):} \qquad  \max _{\{P_k\}_{k=1}^K, \hat{P}_{\rm j}, P_{\rm a}} \left\{\frac{P_{\rm a}}{\varpi \hat{P}_{\rm j} g_{\rm ub} + \sigma_{\rm b}^2} \right\} \label{P1.2:obj} \quad \text{s.t.} \quad  \mathcal{C}1 - \mathcal{C}5.\\[-2pt]
\end{equation*}
To solve \textbf{(P3)}, we also start with the covert constraint $\mathcal{C}1$. The left-hand side of \eqref{eq.covert_cons_reform} can be expressed as $\Phi(p/\Upsilon)$, where $\Phi(\cdot)$ is defined in \eqref{eq.Phi_def}, $p \triangleq P_{\rm a}/\hat{P}_{\rm j}$, and $\Upsilon \triangleq \beta_0^\chi/[\theta \ell_{\rm aw}(\|\mathbf{q} - \mathbf{w}\|^2 + H_{\rm u}^2)]>0$. It can be shown that $\Phi(p/\Upsilon)$ is strictly increasing in $p$ and thus, constraint $\mathcal{C}1$ can be equivalently expressed as\vspace{-3pt}
\begin{equation}\label{eq.C1_reform_P3}
    \mathcal{C}1'': \quad P_{\rm a} \leq \hat{P}_{\rm j} \Phi^{-1}(\varepsilon) \Upsilon. 
\end{equation}

Since $\mathcal{C}1''$, $\mathcal{C}2-\mathcal{C}5$ are all linear constraints, while the objective function of \textbf{(P3)} is a linear-fractional function, \textbf{(P3)} is a linear-fractional program that can be optimally solved by the Charnes-Cooper transformation \cite{Charnes1962Programming}. Specifically, by letting $t\triangleq(\varpi \hat{P}_{\rm j} g_{\rm ub} + \sigma_{\rm b}^2)^{-1} > 0, \tilde{P}_{\rm a}\triangleq P_{\rm a} t, \tilde{P}_{\rm j} \triangleq \hat{P}_{\rm j} t,$ and $\tilde{P}_k \triangleq P_k t,\, \forall k \in \mathcal{K}$, \textbf{(P3)} can be equivalently transformed into the following linear program (LP):\vspace{-3pt}
\begin{align*}
    \textbf{(P3$'$):} \quad &\max _{\{\tilde{P}_k\}_{k=1}^K, \tilde{P}_{\rm j}, \tilde{P}_{\rm a}, t} \quad \tilde{P}_{\rm a} \\
    \text{s.t.} \qquad 
    & \tilde{\mathcal{C}1}: \quad  \tilde{P}_{\rm a} \leq \tilde{P}_{\rm j} \Phi^{-1}(\varepsilon) \Upsilon, \\
    & \tilde{\mathcal{C}2}: \quad  \tilde{P}_k \geq \Xi(\|\mathbf{q} - \mathbf{u}_k\|^2 + H_{\rm u}^2) t,~\forall k \in \mathcal{K}, \\[-1pt]
    & \tilde{\mathcal{C}3}: \quad  \tilde{P}_{\rm j} + \sum_{k=1}^{K} \tilde{P}_k \leq P_{\rm tot} t, \\[-1pt]
    & \tilde{\mathcal{C}4}: \quad  \tilde{P}_{\rm j} \geq 0,~\tilde{P}_k \geq 0,~\forall k\in\mathcal{K}, \\
    & \tilde{\mathcal{C}5}: \quad 0 \leq \tilde{P}_{\rm a} \leq P_{\rm a}^{\max } t, \\
    & \mathcal{C}8: \quad  \varpi g_{\rm ub} \tilde{P}_{\rm j} + t\sigma_{\rm b}^2 = 1, \\
    & {\mathcal{C}}9: \quad t > 0.\\[-16pt]
\end{align*}
Then, \textbf{(P3$'$)} can be efficiently solved using CVX \cite{Boyd_CVX}. Note that the optimal solution of \textbf{(P3)} can be recovered from that of \textbf{(P3$'$)} as $P_{\rm a}^\star = \tilde{P}_{\rm a}^\star / t^\star$, $\hat{P}_{\rm j}^\star = \tilde{P}_{\rm j}^\star / t^\star$, and $P_k^\star = \tilde{P}_k^\star / t^\star,\, \forall k \in \mathcal{K}$.

\subsubsection{Overall Algorithm}
The BCD-based algorithm for solving \textbf{(P1)} is summarized in Algorithm~\ref{Alg.BCD}. 

\begin{algorithm}[t] \caption{Proposed Algorithm for Solving \textbf{(P4)}}
    \small
    \begin{algorithmic}[1]
        \Statex \textbf{Input:} $m, b, \Omega, c, f_{\rm c}, d_{\rm a\chi}, G_{\rm a\chi}, \beta_0^\chi, \beta_0^\kappa$, $\varphi_{\rm min}$, $\sigma_\kappa^2$, $\sigma_{\rm b}^2$, $\sigma_{\rm w}^2$, $H_{\rm min}$, $H_{\rm max}$, $R_{\rm tg}$, $\varepsilon$, $P_{\rm tot}$, $P_{\rm a}^{\rm max}$, $K$, $\mathbf{b}$, $\mathbf{w}$, $\{\mathbf{u}_k\}_{k=1}^K$, $I_{\rm max}$, $\delta$;
        \Statex \textbf{Output:} $\{P_k^\star\}_{k=1}^K, \hat{P}^\star_{\rm j}, P^\star_{\rm a}, \mathbf{q}^\star, H^\star _{\rm u}$.
        \Statex \textbf{Initialize:} Iteration index $i=0$, $\mathbf{q}^{(0)}$, $H_{\rm u}^{(0)}$, and $\lambda^{(0)} = 0$;
        \State Compute $\Phi^{-1}(\varepsilon)$, where $\Phi(\cdot)$ is defined in \eqref{eq.Phi_def};
        \Repeat \Comment{{\it Dinkelbach's algorithm}}
            \State Update $i \gets i + 1$;
            \State Solve \textbf{(P5$^\prime$)} to obtain $\mathbf{q}^{(i)}$ and $H_{\rm u}^{(i)}$;
            \State Update $\lambda^{(i)} \gets f(\mathbf{q}^{(i)}, H_{\rm u}^{(i)})/g(\mathbf{q}^{(i)}, H_{\rm u}^{(i)})$;
        \Until{$|f(\mathbf{q}^{(i)}, H_{\rm u}^{(i)}) - \lambda^{(i-1)} g(\mathbf{q}^{(i)}, H_{\rm u}^{(i)})| < \delta$ or $i>I_{\rm max}$};
        \State Set $\mathbf{q}^\star \gets \mathbf{q}^{(i)}$, $H^\star _{\rm u} \gets H_{\rm u}^{(i)}$, $P_k^\star \gets \Xi[\|\mathbf{q}^\star - \mathbf{u}_k\|^2 + (H^\star_{\rm u})^2]$, $\forall k \in \mathcal{K}$,  $\hat{P}^\star_{\rm j} \gets P_{\rm tot} - \sum_{k=1}^{K}P_k^\star$, and $P_{\rm a}^\star \gets \min\{\lambda^{(i)},P_{\rm a}^{\max}\}$;
    \end{algorithmic}
    \label{Alg.Dinkelbach}
\end{algorithm}

\begin{figure*}[t]
    \centering
    \begin{minipage}[t]{0.3\textwidth}
        \centering
        \includegraphics[width=0.99\linewidth]{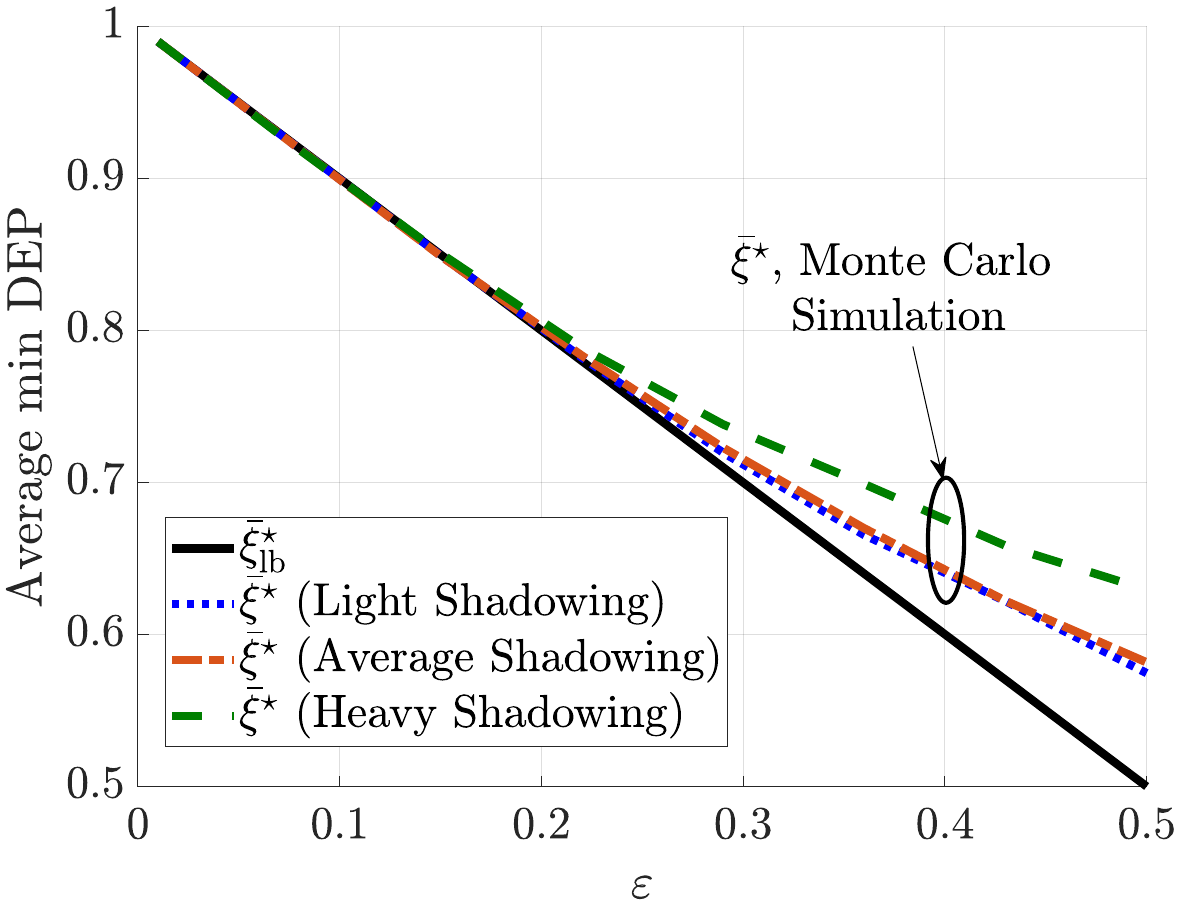}
        \caption{Comparison between average minimum DEP (via MC simulation) and proposed lower bound under different shadowing levels.}
        \label{Fig:DEP_Tightness}
    \end{minipage}
    \hfill
    \begin{minipage}[t]{0.3\textwidth}
        \centering
        \includegraphics[width=0.99\linewidth]{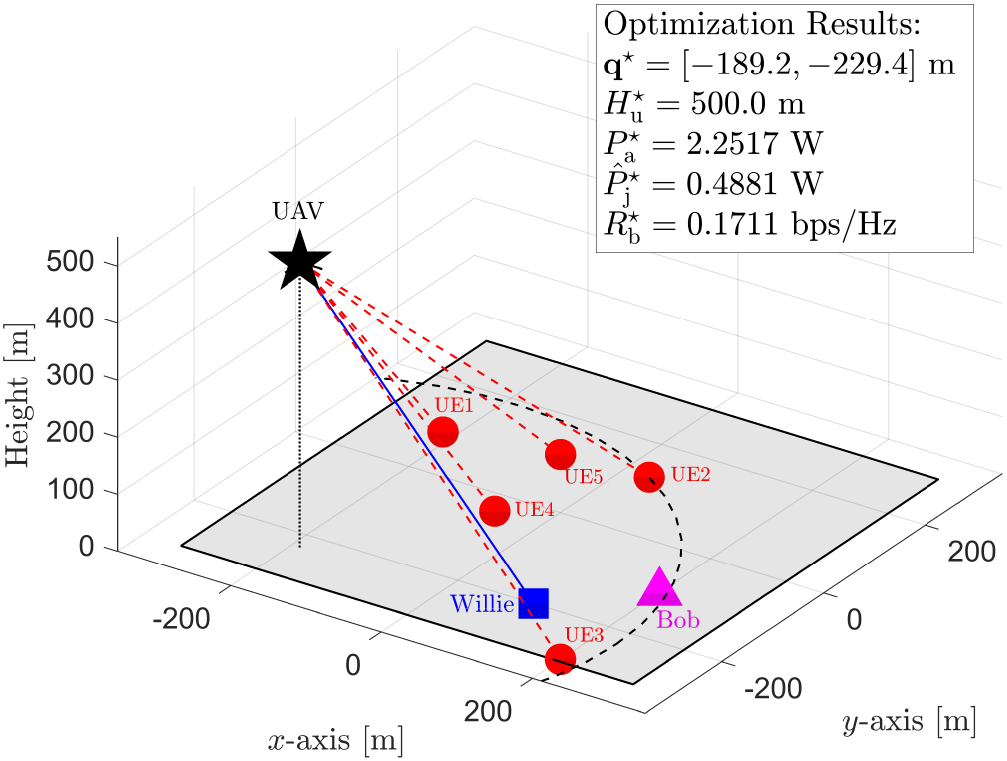}
        \caption{3D visualization of optimal UAV placement and system configuration under imperfect interference cancellation ($\varpi = 0.1$).}
        \label{Fig:3DLocation_Imperfect}
    \end{minipage}
    \hfill
    \begin{minipage}[t]{0.3\textwidth}
        \centering
        \includegraphics[width=0.99\linewidth]{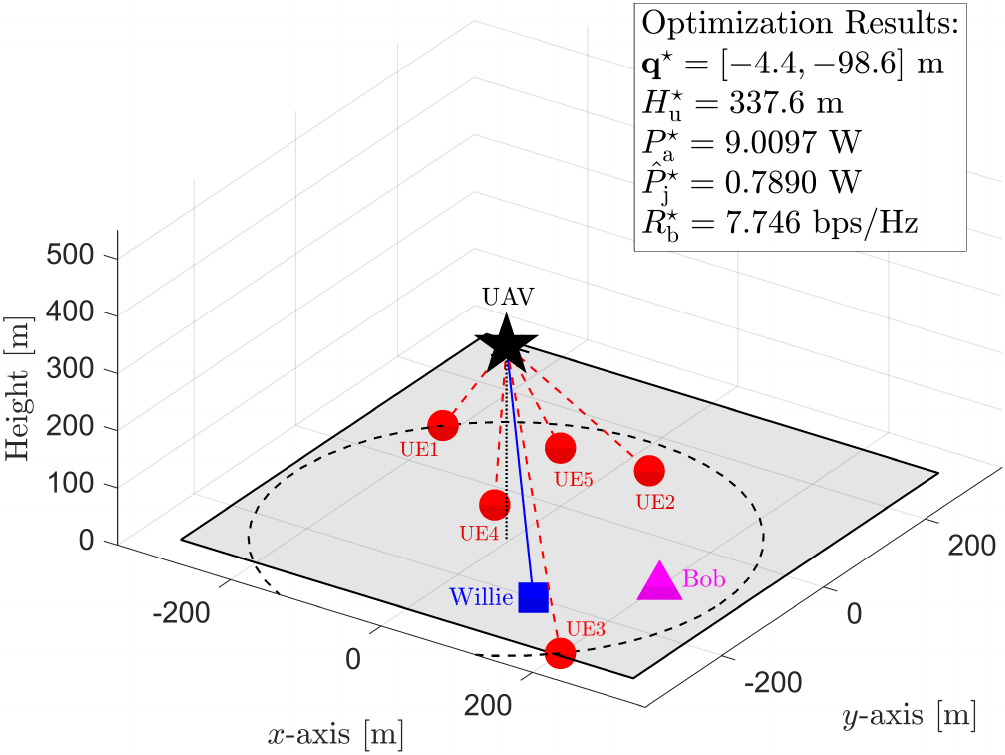}
        \caption{3D visualization of optimal UAV placement and system configuration under perfect interference cancellation ($\varpi =0$).}
        \label{Fig:3DLocation_Perfect}
    \end{minipage}
    \vspace{-10pt}
\end{figure*}

\subsection{Perfect Interference Cancellation Case}
When Bob can perfectly cancel the UAV's jamming signal, i.e., $\varpi = 0$, the covert rate expression in \eqref{eq.R_b} reduces to $R_{\rm b}~=~\log_2(1 + {P_{\rm a}g_{\rm ab}}/{\sigma_{\rm b}^2})$, which is a strictly increasing function of variable  $P_{\rm a}$ only. Problem \textbf{(P1)} thus simplifies to\vspace{-1pt}
\begin{equation*}
    \textbf{(P4):} \qquad 
    \max _{\{P_k\}_{k=1}^K, \hat{P}_{\rm j}, P_{\rm a}, \mathbf{q}, H_{\rm u}} P_{\rm a} \quad 
    \text{s.t.} \quad  \mathcal{C}1 - \mathcal{C}7.
\end{equation*}
\noindent To solve \textbf{(P4)}, we temporarily ignore constraint $\mathcal{C}5$, as it does not affect the optimal solution if $P_{\rm a}^{\max}$ is sufficiently large. 
Now, the covert constraint $\mathcal{C}1$ can be expressed as in \eqref{eq.C1_reform_P3}. Meanwhile, constraint $\mathcal{C}2$ is equivalent to $P_k \geq \Xi(\|\mathbf{q} - \mathbf{u}_k\|^2 + H_{\rm u}^2)$,  $\forall k \in \mathcal{K}$, which, combined with constraint $\mathcal{C}3$, yields\vspace{-3pt}
\begin{equation}
    \hat{P}_{\rm j} \overset{(i)}{\leq} P_{\text{tot}} - \sum_{k=1}^{K} P_k \overset{(ii)}{\leq} P_{\text{tot}} - \Xi \sum_{k=1}^{K} (\|\mathbf{q} - \mathbf{u}_k\|^2 + H_{\rm u}^2). 
    \label{eq.Pj_upperbound}
\end{equation}
Note that both equalities in $(i)$ and $(ii)$ hold when the UAV allocates the minimum required power to each UE to satisfy constraint $\mathcal{C}2$, and uses all the remaining power for jamming. From \eqref{eq.C1_reform_P3} and \eqref{eq.Pj_upperbound}, we have an upper bound for $P_{\rm a}$ as follows:\vspace{-3pt}
\begin{equation}
    P_{\rm a} \leq \left(P_{\text{tot}} - \Xi \sum_{k=1}^{K} (\|\mathbf{q} - \mathbf{u}_k\|^2 + H_{\rm u}^2)\right) \Phi^{-1}(\varepsilon)\Upsilon.
    \label{eq.Pa_upperbound}
\end{equation}
Then, to maximize $P_{\rm a}$, it is desired to make the upper bound in \eqref{eq.Pa_upperbound} as large as possible by optimizing the UAV's placement $\mathbf{q}$ and $H_{\rm u}$. This leads to the following optimization problem:\vspace{-1pt}
\begin{align*}
    \textbf{(P5):} \quad 
    \max_{\mathbf{q}, H_{\rm u}} &\frac{( P_{\rm tot}-\Xi\sum_{k=1}^K  (\|\mathbf{q}-\mathbf{u}_k\|^2 + H_{\rm u}^2))\Phi^{-1}(\varepsilon)\beta_0^\chi}{(\|\mathbf{q}-\mathbf{w}\|^2+H_{\rm u}^2)\theta \ell_{\rm aw} } \hspace{5pt}\\
    &\text{s.t.} \qquad \mathcal{C}6, \mathcal{C}7.
\end{align*}
Problem \textbf{(P5)} is a fractional program with convex constraints and concave-convex fractional objective function. This allows us to apply the Dinkelbach's algorithm \cite{Dinkelbach1967On} to solve it optimally. In particular, we can iteratively solve the following convex optimization problem by CVX until convergence:\vspace{-1pt} 
\begin{equation*}
    \textbf{(P5$^\prime$):} \qquad  
    \max_{\mathbf{q}, H_{\rm u}} \{f(\mathbf{q}, H_{\rm u}) - \lambda g(\mathbf{q}, H_{\rm u})\}  \quad  \text{s.t.} \quad  \mathcal{C}6, \mathcal{C}7.\\[-1pt]
\end{equation*}
Here, $f(\mathbf{q}, H_{\rm u})$ and $g(\mathbf{q}, H_{\rm u})$ denote the numerator and denominator of the objective function in \textbf{(P5)}, respectively, and $\lambda$ is a scalar that is updated in each iteration to drive the objective function of \textbf{(P5$^\prime$)} to zero.

The detailed algorithm for solving \textbf{(P4)} is summarized in Algorithm~\ref{Alg.Dinkelbach}. Note that if $P_{\rm a}^\star$ (i.e., the converged $\lambda^{(i)}$ from the Dinkelbach's algorithm) exceeds $P_{\rm a}^{\max}$, we set $P_{\rm a}^\star = P_{\rm a}^{\max}$ to ensure constraint $\mathcal{C}5$ is satisfied, i.e., $P_{\rm a}^\star \gets \min\{\lambda^{(i)},P_{\rm a}^{\max}\}$.



\section{Numerical Results}\vspace{-2pt} \label{sec.NumerResult}
In this section, we present numerical results to evaluate the proposed algorithms' performance. For simulation settings, we consider $K=5$ UEs uniformly distributed over a square area $[-A/2,A/2]\times[-A/2,A/2]$, with $A=600$~m. Bob and Willie are located at $\mathbf{b} = [A/3,-A/6]^T$ and $\mathbf{w} = [A/6,-A/3]^T$, respectively. We assume light shadowing, and other parameters are set as follows:  $f_{\rm c} = 2$~GHz, $d_{a\chi} = 500$~km, $G_{a\chi}=30$~dBi, $\beta_0^\chi = -38.5$~dB, $\beta_0^\kappa = -60$~dB, $\varphi_{\rm min} = 50^\circ$, $\sigma_\kappa^2 = -114$~dBm, $\sigma_{\rm b}^2 = \sigma_{\rm w}^2 = -104$~dBm, $H_{\rm min} = 50$~m, $H_{\rm max}=500$~m, $R_{\rm tg}= 6$~bps/Hz, $\varepsilon = 0.01$, $P_{\rm tot} = 1$~W, $P_{\rm a}^{\rm max} = 10$~W, $I_{\rm max} = 50$, and $\delta = 10^{-6}$. Besides, $10^4$ channel realizations are generated via Monte Carlo (MC) simulations to compute the average minimum DEP and covert rate. 

Fig.~\ref{Fig:DEP_Tightness} compares the average minimum DEP $\bar{\xi}^\star$ in \eqref{eq.avg_min_DEP}, obtained via MC simulations, with the proposed lower bound $\bar{\xi}^\star_{\rm lb}$ in \eqref{eq.avg_min_DEP_lb} for varying values of $\varepsilon$ under various shadowing levels. It is observed that $\bar{\xi}^\star_{\rm lb}$ closely matches the exact value $\bar{\xi}^\star$, especially when the covert requirement is stringent (i.e., small $\varepsilon$) which is the regime of practical interest. 
This demonstrates that $\bar{\xi}^\star_{\rm lb}$ provides an accurate analytical approximation of $\bar{\xi}^\star$, making it a tractable tool for covertness analysis and yielding high-quality solutions to the original optimization problem.

Figs.~\ref{Fig:3DLocation_Imperfect} and \ref{Fig:3DLocation_Perfect} illustrate the optimal UAV placement and system configuration under imperfect ($\varpi = 0.1$) and perfect ($\varpi = 0$) interference cancellation, obtained by Algorithm~\ref{Alg.BCD} and Algorithm~\ref{Alg.Dinkelbach}, respectively. 
Comparing the two scenarios, we observe that when $\varpi = 0.1$, the UAV tends to stay farther from Bob (i.e., $\mathbf{q}^{\star} = [-189.2, -229.4]^T$~m with maximum altitude $H_{\rm u}^{\star} = 500$~m) to mitigate unintended interference to Bob. In contrast, for $\varpi = 0$, the UAV is positioned closer to the area center (i.e., $\mathbf{q}^{\star} = [-4.4, -98.6]^T$~m with a lower altitude $H_{\rm u}^{\star} = 337.6$~m), thereby enhancing user service while maintaining strong jamming toward Willie.

It is also observed that for $\varpi = 0.1$, the allocated jamming power is limited to $\hat{P}_{\rm j}^{\star} = 0.4881$~W to avoid excessive interference to Bob. Although Alice could potentially increase her transmit power to compensate for the residual interference at Bob, she is constrained to a low transmit power (i.e., $P_{\rm a}^{\star} = 2.2517$~W) to maintain the system's covertness. As a result, the achievable covert rate is only $R_{\rm b}^{\star} = 0.1711$ bps/Hz.
Conversely, for $\varpi = 0$, the UAV can allocate more jamming power (i.e., $\hat{P}_{\rm j}^{\star} = 0.7890$~W) to increase Willie's confusion without harming Bob. This allows Alice to transmit at a higher power (i.e., $P_{\rm a}^{\star} = 9.0097$~W), resulting in a dramatically improved covert rate of $R_{\rm b}^{\star} = 7.746$ bps/Hz.

\section{Conclusion} \label{sec.Concl}\vspace{-2pt}


In this work, we have investigated the problem of UAV-assisted satellite covert communication, where a UAV serves its own network while simultaneously jamming a ground warden to enable covert communication between a satellite and a ground user. 
In future work, we will explore equipping the UAV with multiple antennas and designing beamforming strategies to enhance the covert communication performance.


\ifCLASSOPTIONcaptionsoff
  \newpage
\fi

\bibliographystyle{IEEEtran}
\bibliography{Refs}

\end{document}